\documentclass[notitlepage,a4paper,aps]{revtex4}
\usepackage[usenames,dvipsnames]{xcolor}
\usepackage{amssymb}
\usepackage{amsmath}
\usepackage{amssymb}
\usepackage{amsfonts}
\usepackage{amsthm}
\usepackage{bm}
\usepackage{latexsym}
\usepackage{hyperref}
\usepackage{cleveref}
\usepackage{graphicx}
\usepackage{tikz}
\usetikzlibrary{arrows}

\hypersetup{
    colorlinks,
    citecolor=blue,
    linkcolor=blue,     urlcolor=blue
}

\begin{document}

\title{A spectral unaveraged algorithm for free electron laser simulations}
\author{I.A. Andriyash}
\affiliation{Laboratoire d'Optique Appliqu\'ee, ENSTA-ParisTech, CNRS, Ecole Polytechnique, UMR 7639, 91761 Palaiseau,
France}
\affiliation{P. N. Lebedev Physics Institute, Russian Academy of Sciences, Moscow 119991, Russia}
\email{igor.andriyash@ensta-paristech.fr}
\author{R. Lehe}
\author{V. Malka}
\affiliation{Laboratoire d'Optique Appliqu\'ee, ENSTA-ParisTech, CNRS, Ecole Polytechnique, UMR 7639, 91761 Palaiseau,
France}

\begin{abstract}

We propose and discuss a numerical method to model electromagnetic emission from the oscillating relativistic charged
particles and its coherent amplification. The developed technique is well suited for free electron laser simulations,
but it may also be useful for a wider range of physical problems involving resonant field-particles interactions. The
algorithm integrates the unaveraged coupled equations for the particles and the electromagnetic fields in a discrete
spectral domain. Using this algorithm, it is possible to perform full three-dimensional or axisymmetric simulations of
short-wavelength amplification. In this paper we describe the method, its implementation, and we present examples of
free electron laser simulations comparing the results with the ones provided by commonly known free electron laser
codes.

\end{abstract}

\maketitle

\section{Introduction}\label{intro}

\begin{figure}[h!]\centering
\includegraphics[width=0.45\textwidth]{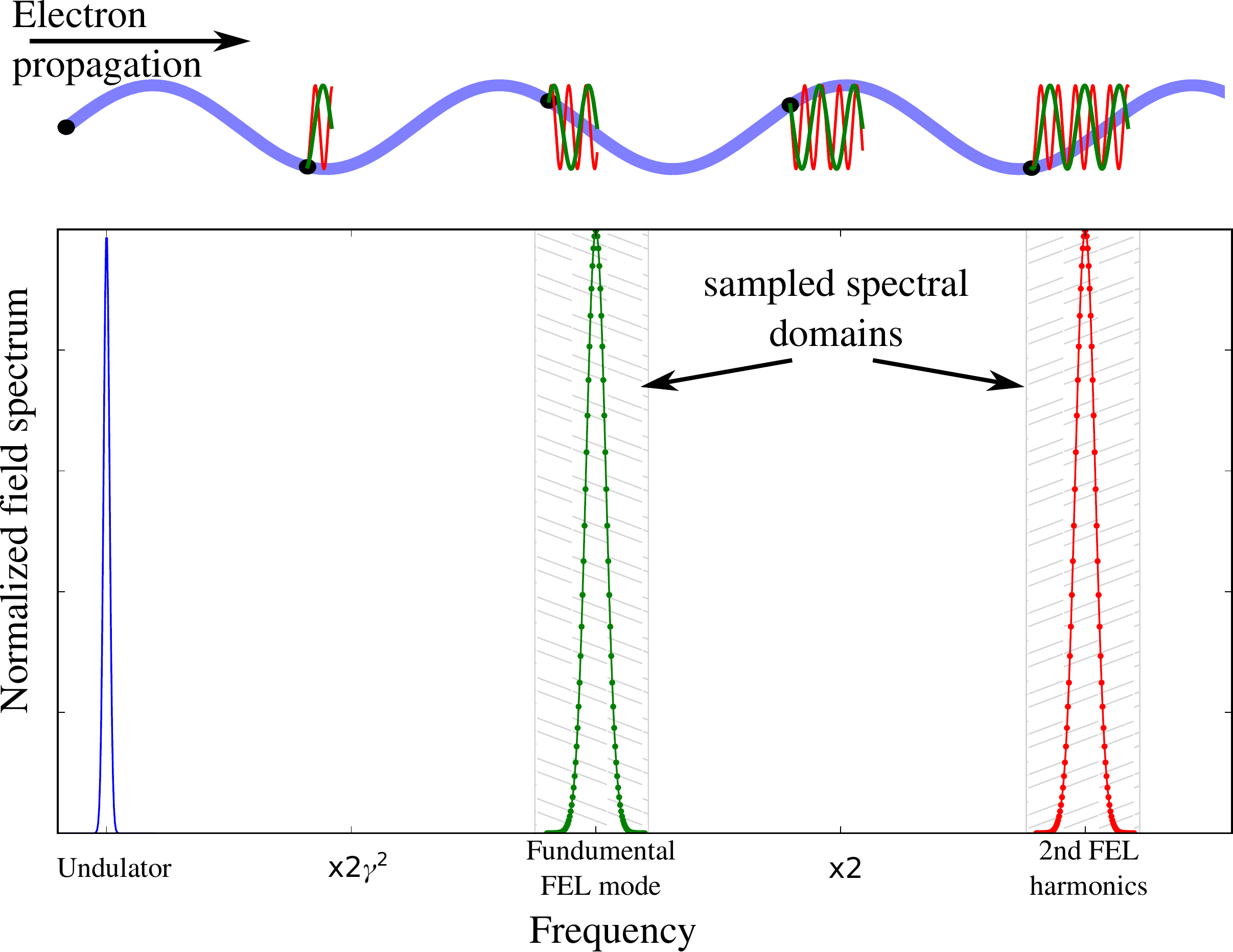}
\caption{Schematic representation of the emission from an oscillating relativistic particle, and of its spectrum.}
\label{spect_scheme}
\end{figure}

When relativistic charged particles propagate through a periodically modulated field, they oscillate and emit a
radiation with a shorter wavelength due to the Doppler effect (see \cref{spect_scheme}). This is the principle of
Synchrotron Radiation (SR) sources, where beams of accelerated electrons are deviated by a wiggling magnetic field in
order to produce bright X-rays. In state-of-the-art sources, the electrons are produced by a linear accelerator and are
wiggled in a series of alternating permanent magnets known as an \textit{undulator}. A particularly interesting case is
that of free electron lasers (FEL), where the SR source operates in a coherent regime.  In these devices, the emitted
X-rays have a resonant interaction with the electrons, and they are efficiently amplified via Stimulated Compton
Scattering (see \cite{huang:PRSTAB2007,mcneil:NatPhot2010} for recent reviews of FELs). Through this process, X-ray free
electron lasers (XFELs) can currently reach the multi-GW level in peak-power, at the angstr\"om wavelengths
\cite{emma:Nat2010}.

In order to describe the physics of free electron lasers, major theoretical results were obtained both in the quantum
and classical approaches \cite{kim:PRL1986,chin:PRA1992,bonifacio:OptComm1998,saldin:2000}, and appropriate numerical
tools were developed \cite{biedron:NIMA2000}. To simulate the particles-field interaction, one may use an approach
similar to the method adopted in the particle-in-cell (PIC) schemes \cite{langdon:2004}. In this approach, the
Maxwell equations are solved on a spatial grid by the finite difference time-domain (FDTD) methods. The electrons are
represented by charged macro-particles and the particles-radiation interaction is conveyed by interpolation techniques.
In the case of relativistic particles interacting with an X-ray radiation, this requires a grid which accurately
resolves the wavelength of the amplified radiation $\lambda_s$ as well as the electron beam size $l_b$. In XFELs the
ratio $l_b/\lambda_s$ may be as high as $10^5 - 10^6$, and the direct implementation of such methods results in a large
computational load.

In the dedicated FEL codes this problem is overcome by making a few approximations. One approximation, used in the
majority of the mainline FEL codes is the \textit{slowly varying envelope approximation} (SVEA), in which the
electromagnetic field is regarded as a wave, $\mathbf{E}(\mathbf{r},t) = \mathrm{Re}\{\hat{\mathbf{E}}(\mathbf{r},t)
\exp[2\pi i(z-ct)/\lambda_s]\}$, where $\lambda_s$ is the anticipated resonant wavelength \cite{arecchi:QE1965} and
where $\hat{\mathbf{E}}$ is assumed to be slowly-varying in space and time \cite{maroli:OptComm2002}. Often SVEA is
complemented with the \textit{wiggler-averaged} formulation which assumes that the trajectories of oscillating electrons
follow precisely the sinusoidal field of the undulator. This approach allows to consider only slow longitudinal
displacements of electrons with respect to the electromagnetic field without solving the full three-dimensional equation
of motion. Both SVEA and wiggler-averaging provide a great calculation efficiency for the conventional FEL
modeling, however, in the cases, where fast variations of the radiation profile are essential the correct description
may require non-averaged approach \cite{maroli:PRSTAB2011,mcneil:OptComm1999}.

Due to the growing interest for the unaveraged FEL modeling, a number of approaches were developed for one-dimensional
simulations \cite{maroli:PRSTAB2011,bajlekov:PRSTAB2011}, and three-dimensional simulations
\cite{freund:PRE1995,campbell:POP2012}. One approach consists in using the multi-frequency model, which is a
pseudo-spectral time domain (PSTD) method, where the Maxwell equations are integrated in a predefined Fourier domain
\cite{lee:JOSAB2004}. One such model is presented in \cite{piovella:POP1999}, where the author uses
a one-dimensional Fourier decomposition of the fields along the direction of propagation, centered around the resonant
frequency. The orthogonality of the Fourier modes in a finite-length periodic interval allows to write separately the
equations for each mode, and to integrate them in time along with the equations motion for the electrons. Similar
approach, using Fast Fourier Transform (FFT) technique, was developed in \cite{bajlekov:PRSTAB2011} for one-dimensional
unaveraged modeling.

In three-dimensional simulations besides the purely FDTD methods  the spectral techniques are often combined with FDTD
approach, for example in \cite{campbell:POP2012}, where the split-step integration accounts for diffraction via FFT, and
particles-field interaction is resolved with an FDTD Galerkin method. Alternatively, decomposition of the
electromagnetic field into the Gauss-Hermite or Gauss-Laguerre modes is also a popular method, which has been used to
represent the transverse profile of the radiation \cite{tang:QE1985,freund:PRE1995}. The choice of the correct
transverse mode can be advantageous in terms of calculation time and accuracy, and it depends on the symmetry of the
problem and its boundary conditions.

In this paper we describe a general approach for three-dimensional modeling of FEL simulations with a spectral method. A
fully three-dimensional algorithm is based on Cartesian Fourier series, and for the axisymmetric case we use the
cylindrical Fourier-Bessel series. The spectral components of the fields are defined on a discrete grid in Fourier
space, and the particles currents are projected onto this spectral grid. The Maxwell equations are integrated
numerically for each spectral component, and the forces are calculated by summing the Fourier series at the particles
positions. The structure of the spectral domain may be chosen rather flexibly, which provides a simple way to capture
the harmonics of the emitted field (see \cref{spect_scheme}). The required spectral resolution for the field can be
rather modest, which in some cases allows to run the simulations on a desktop computer. In this communication we focus
on the numerical recipe itself and its implementation in the context of the basic FEL interaction.

In the following, we describe the mathematical model (\cref{sec1}) and the main aspects of its implementation in the
code \texttt{PlaRes} (\cref{implement}), and we discuss the choice of the simulation parameters (\cref{mod_param}). In
\cref{sec3} we introduce several FEL simulations which were run with \texttt{PlaRes}, and we compare the results with those
obtained by three commonly-used FEL codes. \Cref{sec4} summarizes the results and gives prospective applications for
\texttt{PlaRes}.

\section{Fields and particles equations}\label{sec1}

The interaction between charged particles and electromagnetic fields can be described by the equations for the vector and
scalar potentials $\mathbf{A}$, $\Phi$ written in the  Lorentz gauge \cite{jackson:1998},
\begin{equation}\label{fld_eq0}
 (\partial_t^2 - c^2\nabla^2)\mathbf{A} = 4\pi c \mathbf{J}\,,\quad 
 (\partial_t^2 - c^2\nabla^2)\Phi = 4\pi c^2 \rho\,,\quad 
  \bm{\nabla}\cdot\mathbf{A} + c^{-1}\partial_t\Phi = 0\,,
 \end{equation} 
along with the equations of motion:
\begin{equation}\label{motion_eq1}
  d_t \mathbf{p}_j = -e \left(\mathbf{E} + c^{-1} \mathbf{v}_j\times
\mathbf{B}\right)\,,\quad d_t \mathbf{r}_j =  \mathbf{v}_j \,,\quad \mathbf{v}_j =\mathbf{p}_j/\left(m^2+
|\mathbf{p}_j|^2/c^2\right)^{1/2}\,. 
\end{equation}
Here $\mathbf{J} = -e \sum_j \mathbf{v}_j \delta(\mathbf{r}_j - \mathbf{r})$ and $\rho = -e \sum_j
\delta(\mathbf{r}_j - \mathbf{r})$ are the electron currents and charge densities, $c$ is the speed of light, and $m$ and
$e$ are the mass of the electron and the elementary charge. \Cref{fld_eq0,motion_eq1} are written in Gaussian units.

The radiation and space-charge fields produced by the particles are defined by \cref{fld_eq0}. These fields act back on
the particles through \cref{motion_eq1} along with the ``external'' fields $\mathbf{E}_\text{ext}$ and
$\mathbf{B}_\text{ext}$ produced by the undulator and the complementary devices (quadruples, chicanes etc):
\begin{equation}\label{fields}
\mathbf{E} = \mathbf{E}_\text{ext} -c^{-1} \partial_t \mathbf{A} - \bm{\nabla}\Phi\,,\quad 
\mathbf{B} = \mathbf{B}_\text{ext} + \bm{\nabla}\times \mathbf{A}\,. 
\end{equation}
In order to simulate the looped interaction of the particles and the fields, \cref{fld_eq0,motion_eq1,fields} have to be
self-consistently integrated in time.

The first two equations of the system \cref{fld_eq0} are used to calculate the vector and scalar potentials, and they
can be integrated by a spectral time-domain method. In the following text we will work in dimensionless units. The
potentials are normalized as $\mathbf{a} = e\mathbf{A}/m_ec^2$ and $\phi = e\Phi/m_ec^2$, and the field components are
normalized as $\bm{\varepsilon} = e\mathbf{E}/(mc^2k_u)$ and $\mathbf{b} = e\mathbf{B}/(mc^2k_u)$. The velocities,
spatial coordinates and time are in units of $c$, $k_u^{-1}$ and $\omega_u^{-1}$ respectively, where
$k_u=2\pi/\lambda_u$ is the wavevector of the undulator with a period $\lambda_u$ and $\omega_u = k_u c$.

\subsection{Cartesian Fourier series}\label{sec1:1}

Let us demonstrate the calculation of the radiation field in \cref{fld_eq0} using the Fourier decomposition technique.
For this we consider only the component of $\mathbf{a}$ which is oriented along the electron oscillations. The other
components of $\mathbf{a}$, as well as $\phi$ can be calculated with a similar approach. Using the Fourier theorem and
considering a periodic simulation domain of size $(L_x\,,L_y\,,L_z)$, one may write the fields as a Fourier series:
\begin{equation}\label{spect_decomp}
\mathbf{a}(t,\mathbf{r}) = \Re\left[\sum_{\mathbf{k}\in\mathcal{K}} \mathbf{\tilde{a}}(t, \mathbf{k})
\mathrm{e}^{ -\mathrm{i} \mathbf{k\cdot r}}\right]\,,
\end{equation}
where the wavevector $\mathbf{k}$ is normalized to $k_u$ and belongs to a Cartesian grid $\mathcal{K}$ in Fourier
space. The resolution of $\mathcal{K}$ is defined by the size of the simulation domain as:
\begin{equation}\label{period_cond}
\Delta k_x = 2\pi /L_x\,, \Delta k_y = 2\pi /L_y\,, \Delta k_z = 2\pi /L_z\,.
\end{equation}
The complex values of $\mathbf{\tilde{a}}(t, \mathbf{k})$ define the phases and amplitudes of the Fourier modes at time
$t$. Substituting the decomposition \cref{spect_decomp} into the electromagnetic equation \cref{fields} we obtain:
\begin{equation}\label{fld_eq3}
 \Re\left[\sum_{\mathbf{k}\in\mathcal{K}}\mathrm{e}^{ -\mathrm{i}\mathbf{k\cdot r}}
 \left(\partial_t^2 + |\mathbf{k}|^2\right) \mathbf{\tilde{a}}(t, \mathbf{k})\right] = - \frac{4\pi e^2
 k_u}{m_e c^2} \sum_j \mathbf{v}_{j} \delta(\mathbf{r}_j - \mathbf{r})\,.
\end{equation}

In free electron laser interaction, the velocity of the electron beam and the group velocity of the radiation are
typically very close to $c$, and we may consider the solution to be of the form $\mathbf{\tilde{a}} =
\mathbf{\hat{a}}e^{i |\mathbf{k}| t}$. Using the orthogonality of the Fourier modes in \cref{fld_eq3}, one can multiply
this equation by a factor
$\mathrm{e}^{\mathrm{i}\mathbf{k}_m\mathbf{\cdot r}}$, and integrate it over the simulation domain. This results in a
separate equation for each mode:
\begin{equation}\label{fld_eq4_nosvea}
 \partial_t^2 \mathbf{\hat{a}}(t, \mathbf{k}_m) +  2 \mathrm{i}\,|\mathbf{k}_m|\,\partial_t \mathbf{\hat{a}}(t,
\mathbf{k}_m) = - \frac{8\pi e^2}{m_e \omega_u^2 V}
\sum_j \mathbf{v}_{j} \mathrm{e}^{\mathrm{i}(\mathbf{k}_m\mathbf{\cdot r_j} - |\mathbf{k}_m| t)}\,,
\end{equation}
where $m$ is the index of the mode, and $V = L_x L_y L_z$ is the volume of the simulation domain.

Note that, considering $\mathbf{\tilde{a}} = \mathbf{\hat{a}}e^{i |\mathbf{k}| t}$, we are generally neglecting the
backscattered wave, since its correct description would require a very short integration time-step to resolve the
oscillations of $\mathbf{\hat{a}}$ with a frequency $\sim|2\mathbf{k}|$. 

\subsection{Cylindrical Fourier-Bessel series}\label{sec1:2}

In three-dimensional simulations the algorithm based on the Cartesian Fourier series is able to model the amplification
of a radiation having a complex transverse profile (see \cref{mod_param}). In this case, even for a moderate spectral
resolution, the three-dimensional spectral domain may be very large, and can result in a considerable computational
load. Fortunately, in FEL simulations it is often possible to assume that the produced radiation has a symmetry with
respect to the axis of propagation of the electrons. 

To demonstrate how axial symmetry can be used in the calculations, let us write the field equation (\ref{fld_eq0}) in
the cylindrical coordinates $(z\,,r\,,\theta)$:
\begin{equation}\label{fld_eq6}
 \partial_t^2 \mathbf{a}  - \partial_z^2 \mathbf{a} - \frac{1}{r} \partial_\theta^2 \mathbf{a} - \frac{1}{r}\partial_r(
r \partial_r \mathbf{a}) = - \frac{4\pi e^2k_u}{m_e c^2} \sum_j \mathbf{v}_{j} \delta(\mathbf{r}_j - \mathbf{r})\,.
\end{equation}
The solution to this equation may be constructed with help of cylindrical harmonics, i.e. by decomposing the field in a
Fourier-Bessel series \cite{kaplan:2002}:
\begin{equation}\label{spect_decomp2}
\mathbf{a}(t,\mathbf{r}) = \Re\left[\sum_\alpha \sum_{\mathbf{k}\in\mathcal{K}_\alpha} \mathbf{\tilde{a}}_\alpha(t,
\mathbf{k}) \mathrm{e}^{ -\mathrm{i} k_z z} J_\alpha(k_r r) \cos (\alpha\theta + \Theta_\alpha) \right]\,,
\end{equation}
where $\alpha$ is an integer number, and $\Theta_\alpha$ is an angular phase. The discrete transverse
wavevectors $k_r$ in the Fourier grid $\mathcal{K}_\alpha$
are not necessarily regularly spaced. In order to determine these discrete values of $k_r$ we impose Dirichlet boundary
condition: $\mathbf{a}_\perp(r=R)\equiv 0$, where $R$ is the outer boundary of simulation domain $r\in(0,R)$. This
imposes $k_{r\,j} = u_j/R$ where the $u_j$ are the roots of the equation $J_\alpha(u_j)=0$, and moreover with these
conditions the chosen basis is orthogonal. Using this property of orthogonality, we can find the equation for any given
mode:
\begin{equation}\label{fld_eq7_nosvea}
 \partial_t \mathbf{\hat{a}}_\alpha(t, \mathbf{k}) + 2\mathrm{i}\,|\mathbf{k}|\,\partial_t \mathbf{\hat{a}}_\alpha(t,
\mathbf{k}) = - \frac{8\pi e^2}{m_e \omega_u^2 V (1+\delta_{0,\alpha}) J_{\alpha+1}(k_r R)^2} \sum_j \mathbf{v}_{j}
\mathrm{e}^{\mathrm{i}(k_z z - |\mathbf{k}| t)} \cos(\alpha\theta+ \Theta_\alpha) J_\alpha(k_r r_j)\,,
\end{equation}
where the volume of the cylindrical simulation domain is $V=\pi R^2 L_z$.

The most of the typical situations can be modeled with only first two the terms in the $\alpha$ sum of
\cref{spect_decomp2}. The term $\alpha=0$ defines fully axisymmetric field profile, which does not depend on the angle
$\theta$, while $\alpha=1$ corresponds to the antisymmetric field profile. The phase $\Theta_1$ ($\Theta_0$ may always
be assumed to be 0) allows to choose the antisymmetry of the modeled field. In the FELs this case is typical
for the even harmonics emitted in the linear undulator, which are generated off-axis \cite{huang:PRSTAB2007}.

\subsection{Lorentz force and particle motion}\label{sec1:4}

To model the particle motion, the electric and magnetic fields, $\bm{\varepsilon}$ and $\mathbf{b}$, have to be
calculated from the potentials as $ - \partial_t \mathbf{a} - \mathbf{\nabla} \phi$ and $ \mathbf{\nabla}\times
\mathbf{a}$ respectively. In the case of the Cartesian Fourier decomposition \cref{spect_decomp} we find the fields
acting on the particles as:
\begin{subequations}\label{fld_eq5}
 \begin{equation}\label{fld_eq5:1}
  \bm{\varepsilon}(\mathbf{r}_j, t) = \bm{\varepsilon}_\text{ext}(\mathbf{r}_j,t)  -
   \Re\left[\sum_{\mathbf{k}\in\mathcal{K}} \left[ \partial_t\mathbf{\hat{a}} + \mathrm{i}\,
   |\mathbf{k}|\,\mathbf{\hat{a}}(t, \mathbf{k}) - i \mathbf{k}\,\hat{\phi}(t, \mathbf{k}) \right] \mathrm{e}^{
i|\mathbf{k}| t-\mathrm{i} \mathbf{k\cdot
   r}_j}\right]\,,
 \end{equation}
 \begin{equation}\label{fld_eq5:2}
  \mathbf{b}(\mathbf{r}_j, t) = \mathbf{b}_\text{ext}(\mathbf{r}_j,t)  - \Re\left[\sum_{\mathbf{k}\in\mathcal{K}}
\mathrm{i}\, \mathbf{k} \times \mathbf{\hat{a}}(t, \mathbf{k}) \mathrm{e}^{i|\mathbf{k}| t-\mathrm{i} \mathbf{k\cdot
r}_j}\right]\,.
 \end{equation}
\end{subequations}
The fields from \cref{fld_eq5} are then used to model in the equations of motion for
the particles~:
\begin{equation}\label{motion01}
d_t \mathbf{p}_j = -\bm{\varepsilon}(\mathbf{r}_j, t) - \mathbf{v}_j\times\mathbf{b}(\mathbf{r}_j, t)\,, 
\qquad d_t\mathbf{r}_j = \mathbf{v}_j\,.
\end{equation}

The expression equivalent to \cref{diffract,fld_eq5} for the Fourier-Bessel representation \cref{spect_decomp2}
are more cumbersome but can be easily derived writing explicitly the component of differential operators
``${\bm{\nabla}\cdot}$'' and ``${\bm{\nabla}\times}$'', acting on the cylindrical functions in the Cartesian space.

\subsection{Space-charge and diffraction fields}\label{sec1:3}

The role of the space-charge fields associated with the $\hat{\phi}$ is twofold, and one has to distinguish
the effect of the large-scale field of the whole bunch and the small-scale field of the local density modulations. The
large-scale field results in the Coulomb repulsion of the beam in all directions (see \cite{reiser:2008}). On the other
hand, the small-scale fields alter the dispersion properties of the electron beam and can either suppress or enhance the
amplification of radiation. For instance, while the stimulated Compton scattering is damped by these fields, the
stimulated Raman scattering takes advantage of the collective electron oscillations in a plasma wave
\cite{Shiozawa:2004}.

In the present study, we will neglect both these large-scale and short-scale space-charge phenomena, and thus
$\hat{\phi}$ is assumed to be zero in the rest of this article. This also allows to use the resulting gauge condition,
$\bm{\nabla}\cdot\mathbf{A}=0$, to facilitate accounting for the radiation diffraction. Let us write the relation
between field components for the Cartesian Fourier model:
\begin{equation}\label{diffract0}
 k_x \hat{a}_x +  k_y \hat{a}_y +  k_z \hat{a}_z = 0\,.
\end{equation}

In the common case of a linear undulator, the electrons oscillate along the $x$-axis, and they produce a radiation which
is mainly polarized along $x$, i.e. $a_x\gg a_z, a_y$. Moreover, the radiation is emitted mostly along the $z$ axis, and
thus $k_z\gg k_x\,,k_y$. For these reasons, we may neglect the term $k_y \hat{a}_y\ll k_x \hat{a}_x\,,k_z \hat{a}_z$ in
\cref{diffract0}, and find the longitudinal component of the field (which arises as a consequence of diffraction) as:
\begin{equation}\label{diffract}
 \hat{a}_z = - (k_x/k_z) \hat{a}_x\,.
\end{equation}
For two-dimensional electron oscillations, equations (\cref{fld_eq4_nosvea}) or (\ref{fld_eq7_nosvea}) should be solved
for both $a_x\sim a_y$ components, and $a_z$ can be found directly from \cref{diffract0} as $\hat{a}_z = -
(k_x\hat{a}_x + k_y \hat{a}_y)/k_z$. 

\subsection{Slowly varying envelope approximation}\label{sec1:5}

In the \cref{fld_eq4_nosvea,fld_eq7_nosvea,fld_eq5} we operate with the envelope $\mathbf{\hat{a}}$, which describes
the amplitude of electromagnetic field interacting with relativistic particles. In the most of realistic situations,
the oscillations of the X-ray field are much faster than the dynamics of the amplitude -- growth or damping --
and one may assume $\mathbf{\hat{a}}$ varies slowly in time. As a consequence, the second order term $\partial_t^2 
\mathbf{\hat{a}}\ll 2 |\mathbf{k}|\partial_t \mathbf{\hat{a}}$ can be neglected, which corresponds to the SVEA
approach. With this approximation the field equation for the Cartesian solver
(\ref{fld_eq4_nosvea}) will read:
\begin{equation}\label{fld_eq4}
 \mathrm{i}\,|\mathbf{k}_m|\,\partial_t \mathbf{\hat{a}}(t, \mathbf{k}_m) = - \frac{4\pi e^2}{m_e \omega_u^2 V}
\sum_j \mathbf{v}_{j} \mathrm{e}^{\mathrm{i}(\mathbf{k}_m\mathbf{\cdot r_j} - |\mathbf{k}_m| t)}\,,
\end{equation}
the one for the cylindrical model (\ref{fld_eq7_nosvea}):
\begin{equation}\label{fld_eq7}
 \mathrm{i}\,|\mathbf{k}|\,\partial_t \mathbf{\hat{a}}_\alpha(t, \mathbf{k}) = - \frac{4\pi e^2}{m_e \omega_u^2 V
(1+\delta_{0,\alpha}) J_{\alpha+1}(k_r R)^2} \sum_j \mathbf{v}_{j} \mathrm{e}^{\mathrm{i}(k_z z - |\mathbf{k}| t)}
J_\alpha(k_r r_j) \cos(\alpha\theta)\,,
\end{equation}
and the calculation of the field on the particles positions can be simplified by neglecting the slow variations of
potential amplitude $\partial_t\mathbf{\hat{a}}$ in the first equation of \cref{fld_eq5}:
\begin{equation}\label{fld_eq5_svea}
\bm{\varepsilon}(\mathbf{r}_j, t) = \bm{\varepsilon}_\text{ext}(\mathbf{r}_j,t)  - \Re \left[
\sum_{\mathbf{k}\in\mathcal{K}} i|\mathbf{k}|\,\mathbf{\hat{a}}(t, \mathbf{k}) \mathrm{e}^{i|\mathbf{k}| t-\mathrm{i}
\mathbf{k\cdot r}_j}\right]\,.
\end{equation}

The above equations are similar to those of the FDTD codes that are said to use the slowly-varying approximation (SVEA).
However, these FTDT codes usually introduce two distinct approximations : neglecting the second-order derivative
$\partial_t^2 \mathbf{\hat{a}}$, and using a spatial step equal to $\lambda_s$, justified by the fact that the
envelope is varying in space slowly. While the former approximation is valid when $\partial_t^2 \mathbf{\hat{a}} \ll
k_s \partial_t \mathbf{\hat{a}}$, the latter turns out to have a narrower range of validity. Indeed, recent studies
\cite{maroli:PRSTAB2011,bajlekov:PRSTAB2011} showed that, when solving the first order-field equations on a grid having
a spatial resolution lower than $\lambda_s$, one can correctly model the situations where the radiation profile changes
rapidly, and where the latter approximation is not applicable.

Since our algorithm uses the first-order equation but does not introduce this coarse spatial discretization of
$\lambda_s$, it may have a wider range of applicability than standard FTDT codes. This is confirmed in the
\cref{sec3.1}, where we show that the second-order equation eqs. (\ref{fld_eq4_nosvea},\,\ref{fld_eq7_nosvea}) and
first-order eqs. (\ref{fld_eq4},\,\ref{fld_eq7}) lead to very similar result, even when the radiation profile
dynamics is very rapid.

\section{Numerical integration scheme and its implementation}\label{implement}

The methods described in the previous section may describe the X-ray field amplification with account for the
three-dimensional effects (radiation diffraction and electron divergence) at a moderate computational cost. In the
simulations, the field and particle equations have to be numerically integrated over time. To integrate the first order
differential equations we choose a leapfrog algorithm, since it requires a single computation of the fields at each
iteration, while retaining second-order accuracy. In this approach, the values of the forces and the particles positions
are defined at integer times ($n\Delta t$), while the ``generating'' terms (e.g. electron currents) are defined at
half-integer times ($(n+1/2)\Delta t$) -- where $\Delta t$ is the integration timestep. We therefore refer to these
variable with a corresponding superscript of the form $n$ or $n+1/2$ in rest of this section. 

\subsection{Reduced order scheme (SVEA)}\label{implement2:1}

Here let us detail one full cycle of SVEA model integration, i.e. how the variables at iteration $n$ are deduced from
the variables at iteration $n-1$. In order to do so, let us consider that, at the iteration $n-1$, the known variables
are $\mathbf{r}_j^{n-1}$, $\mathbf{\hat{a}}^{n-1}$, $\mathbf{p}_j^{n-1/2}$ and $\gamma_j^{n-1/2} =
(1+|\mathbf{p}_j^{n-1/2}|^2)^{1/2}$. Going from iteration $n-1$ to iteration $n$ involves updating the fields (a), as
well as the particles positions and velocities (b). These successive operations are repeated at each integration
timestep.

\begin{figure}[ht!]\centering
\begin{tikzpicture}
\draw[->, >= stealth] (0,0) -- (8,0) node[below]{$t$};
\draw[fill=black] (1,0) circle(0.07cm);
\draw (1,0) node[above]{$\mathbf{r}^{n-1}_j$};
\draw (1,-0) node[below]{$\mathbf{\hat{a}}^{n-1}$};
\draw[fill=gray] (3,0) circle(0.07cm);
\draw (3,0.5) node[above]{(\ref{eq:halfpush_part})$\rightarrow\;${$\mathbf{r}^{n-1/2}_j$}};
\draw (3,0) node[below]{$\mathbf{v}^{n-1/2}_j$};
\draw[fill=black] (5,0) circle(0.07cm);
\draw (5,-0.5) node[below]{(\ref{eq:push_field})$\rightarrow\;${$\mathbf{\hat{a}}^n$}};
\draw[fill=gray] (7,0) circle(0.07cm);
\draw[->,>=triangle 45] (1,0) ..controls (1.5,0.5) and (2.5,0.5) .. (3,0);
\draw[->,>=triangle 45] (1,0) ..controls (2,-1) and (4,-1) .. (5,0);
\draw (0,1) node{\textbf{(a)}};

\begin{scope}[shift={(0,-3)}]
\draw[->, >=stealth] (0,0) -- (8,0) node[below]{$t$};
\draw[fill=black] (1,0) circle(0.07cm);
\draw (1,0) node[above]{$\mathbf{r}^{n-1}_j$};
\draw (1,-0) node[below]{$\mathbf{\hat{a}}^{n-1}$};
\draw[fill=gray] (3,0) circle(0.07cm);
\draw (3,0) node[below]{$\mathbf{p}^{n-1/2}_j$};
\draw[fill=black] (5,0) circle(0.07cm);
\draw (5,-0) node[below]{$\mathbf{\hat{a}}^{n}_j$};
\draw[fill=gray] (7,0) circle(0.07cm);
\draw[->,>=triangle 45] (1,0) ..controls (2,1) and (4,1) .. (5,0);
\draw[->,>=triangle 45] (3,0) ..controls (4,-1) and (6,-1) .. (7,0);
\draw (5,0.5) node[above]{(\ref{eq:push_part})$\rightarrow\;${$\mathbf{r}^{n}_j$}};
\draw (7,-0.5) node[below]{(\ref{eq:Boris})$\rightarrow\;\mathbf{p}^{n+1/2}$};
\draw (0,1) node{\textbf{(b)}};
\end{scope}
\end{tikzpicture}
\caption{Typical leapfrog schemes for the fields (a) and the particles (b)} \label{leaps}
\end{figure}
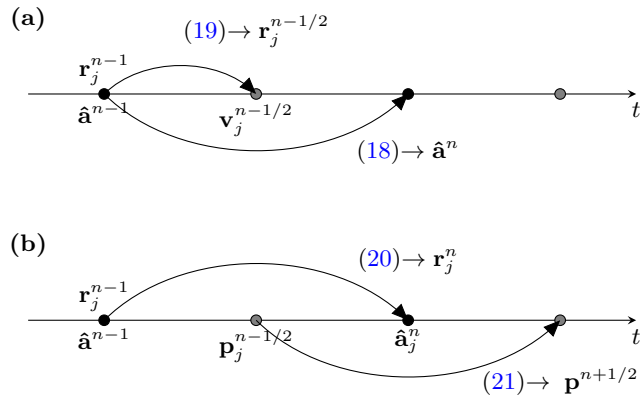

(a) To calculate the electromagnetic field, we consider the leapfrog discretization of \cref{fld_eq4} or
(\ref{fld_eq7}):
\begin{equation}\label{eq:push_field} 
\mathbf{\hat{a}}_{\perp}^{n} = \mathbf{\hat{a}}_{\perp}^{n-1} +i \frac{4\pi e^2 \Delta t}{m_e \omega_u^2 V
|\mathbf{k}|} \sum_j \mathbf{v}^{n-1/2}_{\perp\,j} \exp\left(\mathrm{i}k_z z_j^{n-1/2} -
\mathrm{i}|\mathbf{k}| (n-1/2) \Delta t \right) F(\mathbf{r}_\perp, \mathbf{k}_\perp)\,,
\end{equation}
where $F = \exp\left(\mathrm{i} \mathbf{k}_\perp \cdot\mathbf{r}_{\perp\,j}^{n-1/2}\right)$ for the Cartesian algorithm,
and $F = \cfrac{\cos(\alpha\theta + \Theta_\alpha)}{1+\delta_{0,\alpha}} \cfrac{J_\alpha(k_r r_{\perp\,j}^{n-1/2})}{J_{\alpha+1}(k_r
R)^2}$ for Fourier-Bessel one.

Note that \cref{eq:push_field} requires the variables $\mathbf{r}_j$ and $\mathbf{v}_j = \mathbf{p}_j/\gamma_j$ to be
defined at step $n-1/2$, whereas $\mathbf{r}_j^{n-1/2}$ is not yet known. Therefore, before applying equation
(\ref{eq:push_field}), the algorithm first computes the \emph{auxiliary} variables $\mathbf{r}_j^{n-1/2}$ by taking a
half-timestep: 
\begin{equation}\label{eq:halfpush_part}
\mathbf{r}_j^{n-1/2} = \mathbf{r}_j^{n-1} + \mathbf{v}_j^{n-1/2}  \Delta t/2\,,
\end{equation}
while keeping $\mathbf{r}_j^{n-1}$ in the computers memory.

(b) The particles coordinates are updated by a standard leapfrog operation:
\begin{equation}
\label{eq:push_part}
\mathbf{r}^{n} = \mathbf{r}^{n-1} + \mathbf{v}^{n-1/2} \Delta t\,.
\end{equation}
Calculating the particles velocity first involves computing the electric and magnetic fields at $t=n\Delta t$, by using
equations \cref{fld_eq5_svea,fld_eq5:2} or their counterpart for the Fourier-Bessel method. The values of the
electromagnetic fields are then used in the equation of motion \cref{motion01}. This equation is implemented with the
Boris pusher \cite{boris:PROC1970}, which is an explicit algorithm that extracts $\mathbf{p}^{n+1/2}_j$ from the
implicit discretized equation~: 
\begin{equation}
\label{eq:Boris}
\frac{\mathbf{p}_j^{n+1/2}-\mathbf{p}_j^{n- 1/2}}{\Delta t}  = - \bm{\varepsilon}^n_j - 
\left(\frac{\mathbf{p}_j^{n+1/2}+\mathbf{p}_j^{n-1/2} }{2\bar{\gamma}^n}\right)\times \mathbf{b}^n_j\,, \qquad
\bar{\gamma}^n \equiv \sqrt{1+(\mathbf{p}^{n-1/2} -\bm{\varepsilon}^{n}_j \,\Delta t/2)^2 } \,.
\end{equation}
These operations are summarized in \cref{leaps}.

\subsection{Non-SVEA algorithm}\label{implement2:2}

The procedure of the integration of \cref{fld_eq4_nosvea,fld_eq7_nosvea} is rather similar to the one described above,
and it may be done with help of leapfrog method. For this we first define the time derivative of $\mathbf{\hat{a}}$ as a
new variable $\mathbf{\hat{g}} = \partial_t\mathbf{\hat{a}}$ and write two first order equations:
\begin{subequations}\label{eq:push_field_nosvea}
 \begin{equation}
  \partial_t \mathbf{\hat{g}} +  2 \mathrm{i}\,|\mathbf{k}_m|\,\mathbf{\hat{g}} = S(t,\mathbf{k}_m)\,,
 \end{equation}
 \begin{equation}
  \partial_t\mathbf{\hat{a}} =  \mathbf{\hat{g}}\,,
 \end{equation}
\end{subequations}
where $S(t,k_m)$ is a left hand side of \cref{fld_eq4_nosvea} or \cref{fld_eq7_nosvea}. 

Now, defining $\mathbf{\hat{g}}$ and $\mathbf{\hat{a}}$ at the integer and half-integer time steps respectively, we
can integrate the set \cref{eq:push_field_nosvea} numerically with the discretized equations:
\begin{subequations}\label{eq:push_field_nosvea1}
 \begin{equation}\label{eq:push_field_nosvea1:1}
  \mathbf{\hat{g}}^{n} =  \cfrac{1-\mathrm{i}\,|\mathbf{k}_m|\Delta t}{1+\mathrm{i}\,|\mathbf{k}_m|\Delta
  t}\;\mathbf{\hat{g}}^{n-1}  +  \cfrac{\Delta t}{1+\mathrm{i}\,|\mathbf{k}_m|\Delta t}\; S^{n-1/2}\,,
 \end{equation}
 \begin{equation}\label{eq:push_field_nosvea1:2}
  \mathbf{\hat{a}}^{n+1/2} =  \mathbf{\hat{a}}^{n-1/2} + \Delta t\, \mathbf{\hat{g}}^{n}\,,
 \end{equation}
\end{subequations}
Note that to use the leapfrog scheme for the particle motion, the Lorentz force has to be calculated on the particles
positions at the integer step with help of \cref{fld_eq5}. Therefore, we have to explicitly calculate the value
of $\mathbf{\hat{a}}^{n} = \mathbf{\hat{a}}^{n-1/2} + 0.5\Delta t\, \mathbf{\hat{g}}^{n}$. The other steps are identical
to the ones described in \cref{implement2:1}.

\subsection{Suppression of the shot noise}\label{implement2:3}

Since the real number of electrons in the beam is typically very big (up to $10^{10}$), in numerical simulations, the
electrons are usually replaced by macro-particles. Each macro-particle has the same charge-to-mass ratio as an electron,
but represents a large number of electrons. Typically, in simulations of FEL amplification, a reasonable modeling
accuracy can be reached by using a relatively low number of macro-particles, $\sim 10^3-10^5$. However, the artificially
low number of particles and associated shot noise affect the spontaneous emission generated at the beginning of a FEL
interaction. In practice, this simulated spontaneous emission can be very different from the physical one. In the
case of the self-amplified spontaneous emission process (SASE), the final X-ray signal origins from this spontaneous
noise, and thus the correct modeling of it is very important.

The control of the shot noise in FEL codes have been extensively studied in the last decades
\cite{penman:OptComm1992,fawley:PRSTAB2002,sprangle:PRL1983,mcneil:PRSTAB2003}. The basic principle of this control is
to suppress the shot noise for a given wavelength $\lambda_\mathrm{supp}$, by placing, for each macro-particle, an
identical one with a longitudinal coordinate shifted by $\lambda_\mathrm{supp}/2$, so that it will emit in anti-phase
with its counterpart. The physical shot noise can then be generated via a proper random displacement of these additional
particles. 

For the first test studies of our algorithm, we will use a simple particle loading algorithm, which provides full
noise suppression for a chosen wavelength and its harmonics:
\begin{equation}
 \{z\} = \{z\} \cup \{z+\lambda_\mathrm{supp}/(2\alpha)\} \,,
\end{equation}
where $\{z\}$ is a set of the particles $z$-coordinates, and $\alpha$ is a harmonic number.

\subsection{Code \texttt{PlaRes}}\label{implement2:4}

The algorithms for three-dimensional and axisymmetric simulations presented in this section are implemented in the code
\texttt{PlaRes} (which stands for ``plasma resonator''). The code allows to use and combine the different Maxwell
solvers in the same simulation, where each solver is defined by its own discontinuously composed spectral domain. In
the case of solvers combination, the solvers should be linearly independent (e.g. axially symmetric and antisymmetric), or
operate on non-overlapping spectral domains.

The tool is developed using a hybrid programming framework: the initialization of the simulation, the parallelization,
the runtime management and the data output are performed by a Python object-oriented code, while all the ``expensive''
number-crunching routines are written in Fortran 90 and called from Python via the \textbf{f2py} interface generator
\cite{f2py}. Parallelization is carried out by distributing the particles among the processing units and summing their
contributions to the fields via message passing interface (MPI) (using Python's \textbf{mpi4py} package \cite{mpi4py}).
In this way, the code provides a high flexibility for new external fields setup and for further algorithmic
developments.

\section{Modeling parameters}\label{mod_param}

In order to implement the method of \cref{sec1,implement} in practice, one needs to truncate the series in
\cref{spect_decomp,spect_decomp2}, and to therefore to specify a finite spectral domain in which the studied fields are
localized. The optimal choice for these parameters is based on the physics of synchrotron radiation and free electron
lasers.

The spectral properties of synchrotron radiation are well known, and are described in the literature (see
\cite{hofmann:2004}). The wavelength of the emitted light by a relativistic charge is Doppler-shifted by a factor
$(1-\mathbf{n}\cdot\mathbf{v})^{-1}$, where $\mathbf{n} = \mathbf{k}/|k|$ is a unit vector pointing towards the
observer. For an oscillating relativistic electron with $v_\parallel\simeq 1$ and $v_\perp = \hat{v}_\perp \sin k_u z$,
the emission is produced in the direction of the particle's velocity, and within a cone of angular size
$\sim\gamma_b^{-1}$. For a linearly polarized undulator, the central wavenumber of the emitted light reads:
\begin{equation}\label{res_k} 
k_s = \cfrac{2\gamma_b^2 k_u}{1+K_0^2/2+\gamma_b^2\theta^2}\,, 
\end{equation}
and its bandwidth is determined by the total number of oscillations performed by the particle, $\delta k_z/k_s \simeq
N_\text{osc}^{-1}$. Here $K_0$ is the undulator parameter and $\theta = k_\perp/k_\parallel$ is the angle between the
direction of observation and the electron trajectory. The undulator parameter can be defined as the amplitude of
oscillation of the normalized electron momentum $K_0 = p_\perp/mc = \hat{v}_\perp\gamma_b$. For a magnetic undulator
with a field $B_0$, the undulator parameter reads \cite{huang:PRSTAB2007}: 
\[K_0 = eB_0/(mc^2k_u) = 0.934 \;B_0 [\text{T}]\; \lambda_u [\text{cm}]\,.\]
For large values of $K_0$, the particles trajectories are significantly non-harmonic. In this case, the emitted spectrum
contains harmonics of $k_s$ and becomes broadband.

During resonant amplification, the power of the generated radiation grows exponentially along the undulator
$P_\text{rad} \propto \exp(z/L_{g})$, where $L_{g}$ is the \textit{gain length}. The spectral bandwidth of the amplified
light is typically determined by the spectrum of the light emitted by individual particles over $L_{g}$. In the
simple one-dimensional case of an ideal electrons beam, the gain length and the corresponding growth rate are
determined
by the FEL parameter $\rho$ \cite{bonifacio:OptComm1984}:
\begin{equation}\label{gain1D}
L_{g0} = \frac{1}{2\sqrt{3} \rho k_u}\,, \qquad \rho = \gamma_e^{-1} (n_e/32 n_c)^{1/3} \left([JJ]\, K_0
\right)^{2/3}\,, 
\end{equation}
where $n_e$ is the peak density of the electron bunch and $n_c$ is defined by $n_c \equiv \pi mc^2/e^2\lambda_u^2$. By
definition, $[JJ] \equiv J_0(\xi) - J_1(\xi)$, where $J_0$ and $J_1$ are the Bessel functions and $\xi =
K_0^2/4(2+K_0^2)$. In this formulation, the bandwidth of the amplified light estimates as $\delta k_z/k_s \simeq
\rho$. It is also important to mention, that according to the sampling theorem the spectral width $\delta k_z$ is
directly related to the minimal spatial scale of the modeled radiation. This may become important, when modeling very
short electron currents \cite{maroli:PRSTAB2011}, seeding with a pulse produced by high-harmonics generation
\cite{bajlekov:PRSTAB2011}, etc. In this case, the required spectral bandwidth may be much bigger than the one
estimated in terms of FEL resonance, and should be taken into account.

The transverse size of the spectral region required to describe the shape of the radiation beam and its diffraction can
be deduced from \cref{res_k}, by considering that the amplified light is emitted around the resonant frequency
$k_\parallel\simeq k_s|_{\theta=0}$ with a bandwidth $\delta k_\parallel$. On the whole, the dimensions of the spectral
domain required for an FEL simulation can be approximately defined by the following formulas:
\begin{align}
 & k_\parallel = 2\gamma_z^2 k_u\,,\nonumber\\
 & \delta k_\parallel \gtrsim \rho k_\parallel\,,\\
 & k_\perp\gtrsim \sqrt{2 \delta k_\parallel k_u}\,,\nonumber
\end{align}
where $\gamma_z$ is a Lorentz factor associated with electron longitudinal motion, and in case of a linearly polarized
undulator $\gamma_z = \gamma_b\,(1+K_0^2/2)^{-1/2}$.

It is also important to correctly choose the resolution of this spectral domain, which defines the size of the periodic
simulation domain, according to \cref{period_cond}. The longitudinal size of the simulation domain $L_z$ should be
larger than both the length of the electron beam $l_{\mathrm{beam}}$ and the slippage length of the radiation
$ct_\text{sim}/2\gamma_z^2$. The transverse sizes $L_{x,y}$ or $R$ have to be larger than the radius of the electron
beam $\sigma_\perp$ and the radiation beam size $\sigma_s$. In practice, the structure of the radiation transverse mode
is defined by its diffraction parameter $\eta_d = L_{g0}/(2 k_s \sigma_\perp^2)$ \cite{huang:PRSTAB2007,xie:NIMA2000-2}.
If the diffraction is weak, $\eta_d\ll 1$, the amplification is one-dimensional, and it occurs mainly within the
electron beam i.e. $\sigma_s \simeq \sigma_\perp$. In the opposite case, $\eta_d\gg 1$, diffraction leads to a radiation
transverse size, which is much larger than the electron beam. The parameters of this three-dimensional amplification are
significantly different from \cref{gain1D}, and the radiation transverse size is related to the gain length as:
\begin{equation}\label{gain3D}
\sigma_s \simeq \sqrt{\cfrac{\lambda_s L_{gD}}{2\pi}}\,,\qquad L_{gD} = \cfrac{\lambda_u^2}{8\pi^2\sigma_\perp K_0 [JJ]}
\sqrt{\cfrac{2\gamma_b n_c}{n_e}\, \left(1+K_0^2/2\right)}\,.
\end{equation}

\section{Simulations}\label{sec3}

In order to test the code \texttt{Plares}, we benchmark it against other existing FEL codes. For this we choose the data
of the comparative analysis published by L. Giannessi et al in \cite{giannessi:NIMA2008}. In this article, the authors
compare three well-known codes MEDUSA \cite{freund:PRE1995}, GENESIS 1.3 \cite{reiche:NIMA1999} and PERSEO
\cite{giannessi:PERSEO2006} on a test case which involves a coupling of high-order harmonics in a conventional magnetic
undulator. In the corresponding simulation, the electron beam has a 1 ps flat-profile with a current of 110 Amperes, a
mean energy of 200 MeV, 0.01\% energy spread, a transverse emittance of 1 mm$\cdot$mrad, and a radius of 95.3 $\mu$m.
The electrons travel through a magnetic undulator with a period of 2.8 cm and an undulator parameter $K=1.95$. The FEL
process is \emph{seeded} by a co-propagating 50 fs Gaussian electromagnetic pulse, which is focused to a 183.74~$\mu$m
transverse size, at a distance of 70 cm from the undulator entrance. The wavelength of the seed pulse is matched to the
central wavelength of the undulator radiation $\lambda_s = 265.151$ nm and its peak power is set at 10 kW. The shot
noise in the simulations is completely suppressed. 

In our tests we use the simple model of an ideal linearly polarized undulator, where the magnetic field oscillates with
the longitudinal coordinate, and does not depend on the transverse position. Moreover, no special treatment of
the beam's transverse dynamics is performed, so that, after being initiated with a flat transverse density profile and a
Gaussian distribution in transverse momenta, the beam diverges freely.

In \cite{giannessi:NIMA2008}, the authors perform one- and three-dimensional simulations within the steady-state and
time-dependent approaches. In the steady-state approach, the electron bunch is considered to be infinitely long, and the
radiation amplitude is assumed to be independent of the longitudinal coordinate. This approach is close to the
theoretical description of the FEL process, and, in addition, it allows to reduce the longitudinal size of the
simulation domain to a few radiation wavelengths -- thereby drastically decreasing the computational load.

The time-dependent approach allows to include more phenomena into the simulations and to study the impact of the finite
lengths of the electron and radiation beams. For example, time-dependent simulations are able to model the ``slippage''
effect, i.e. the fact that the emitted radiation progressively overtakes the electron bunch and eventually stops being
amplified. Slippage typically results in significant modulations of the longitudinal profile of the radiation.

\subsection{One-dimensional simulations}\label{sec3.1}

\begin{figure}[h!]\centering
\includegraphics[width=0.9\textwidth]{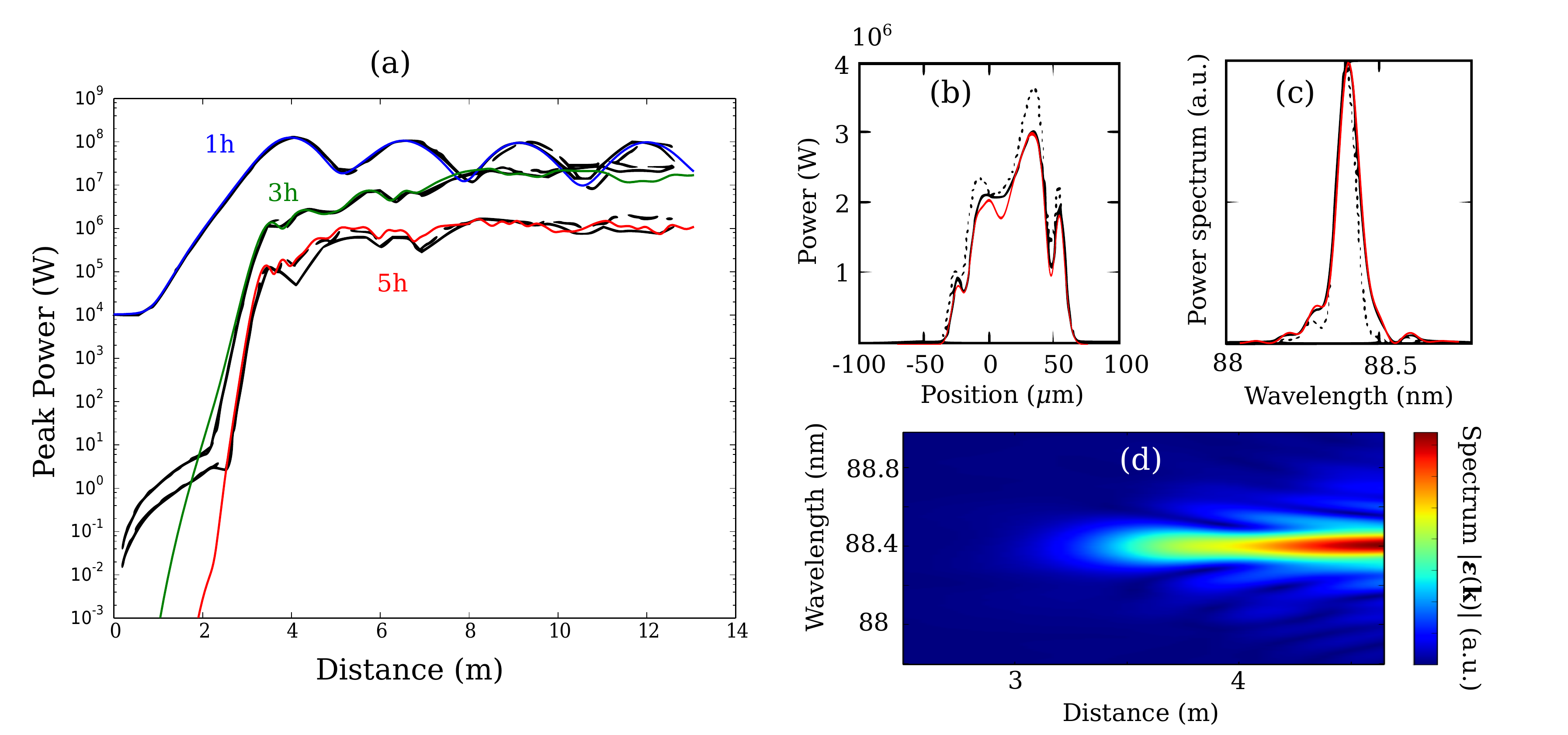}
\caption{(a) Radiation power for the first three odd harmonics, as a function of the propagation distance. (b) and (c) :
spatial (b) and spectral (c) profiles of the third harmonic at $z=4.6$ m. (d) Spectral dynamics of the third harmonics.
In (a,b,c) the results obtained with the code \texttt{PlaRes} 1D (colored curves) is compared with the data produced by
MEDUSA1D (dashed black), PERSEO (solid black) \cite{giannessi:NIMA2008}.}
\label{compare1}
\end{figure}

For the one-dimensional case we calculate the power of the radiation emitted at the first three odd harmonics. In the
steady-state simulations each harmonic is modeled with a single wavenumber, so that the spectral domain is $k_z =
[k_s\,, 3 k_s\,,5 k_s]$. The electron beam is modeled with 2000 macro-particles. In order to capture the fifth harmonic
in the electron oscillations we advance each period of the radiation field with 90 iterations, so that the chosen time
step $\Delta t = \lambda_u/90 c$.

In \cref{compare1}a, we compare the results obtained by \texttt{Plares} (colored curves) with the ones obtained by
MEDUSA1D (black dashed curve) and PERSEO (black solid curve) for steady-state simulations. The three codes agree almost
perfectly in the description of the (seeded) first harmonic, even at the saturation. The third and fifth harmonics are
not seeded, and the comparison reveals that our code develops a lower level of spontaneous noise in the third and
fifth harmonics during the early interaction. The agreement is particularly good with the unaveraged MEDUSA1D
simulations.

In order to study the longitudinal profile of the radiation, we have also run time-dependent simulations. In these
tests, the spectral domain of the fundamental and third FEL harmonics have bandwidths of 1.6\% $k_s$ and 4\% $k_s$
respectively, and the ``time window'' (i.e. length of the periodic simulation domain) is $L_z/c = 1.5$ ps. This results
into a total of 98 spectral modes. As the chosen duration of the electron beam in \cite{giannessi:NIMA2008} is rather long (1 ps),
this simulation requires a big number of macro-particles: $10^5$. In \cref{compare1}b  and \cref{compare1}c we compare
the calculated spectral and spatial profiles of the radiation with those obtained in the MEDUSA1D and PERSEO
simulations. The agreement between the three codes is found to be remarkably good. The dynamics of the spectral
distribution of the third harmonic is shown in \cref{compare1}d. In this figure, the formation, growth and saturation of
the third harmonic can be clearly seen.

\begin{figure}[h!]\centering
\includegraphics[width=0.45\textwidth]{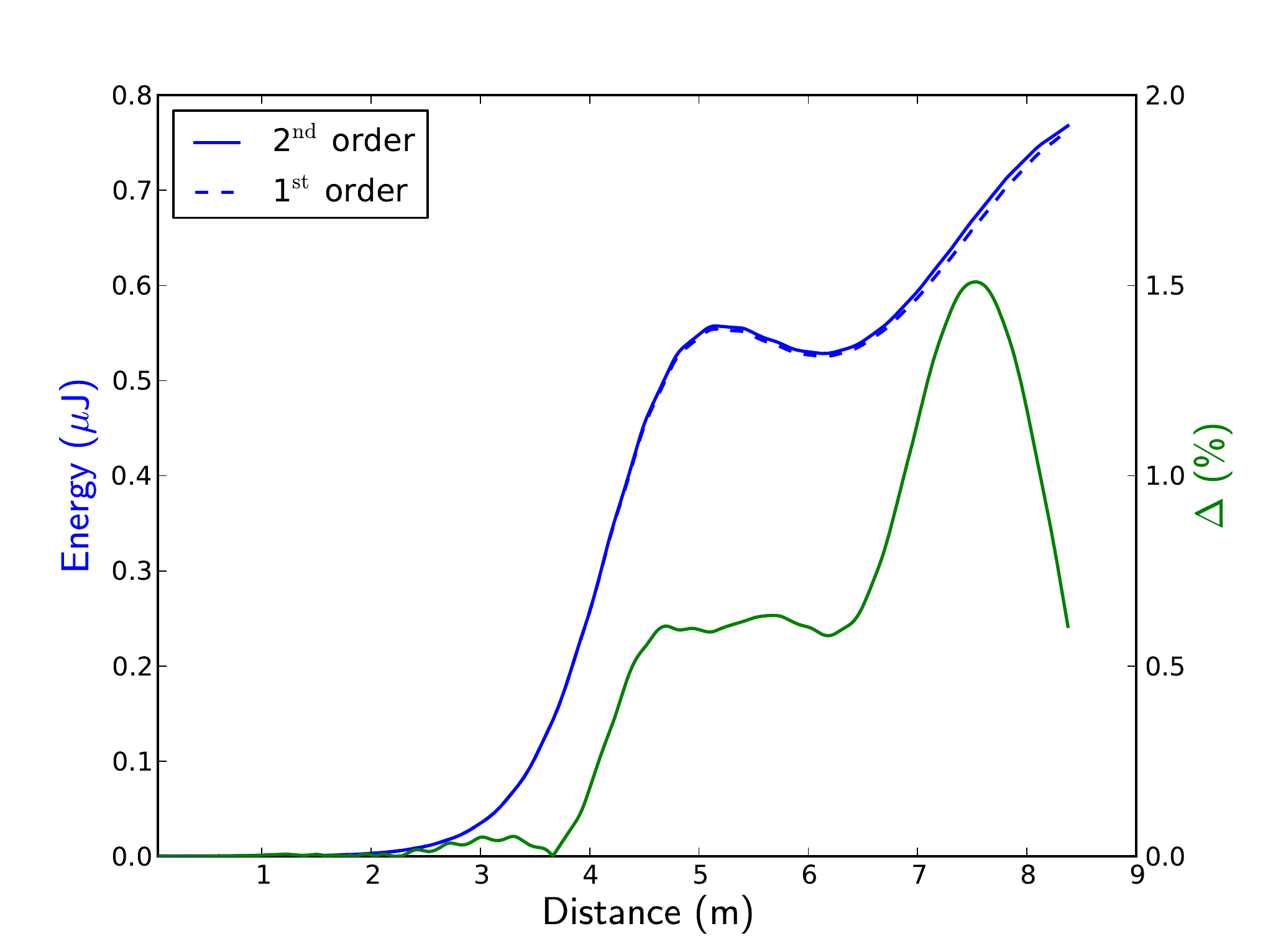}
\caption{Radiation energy in the fundamental harmonic, as a function of the propagation distance for a case of 28 fs
electrons current calculated with non-reduced \cref{implement2:2} (solid blue curve) and reduced \cref{implement2:1}
(dashed blue curve) algorithms. Dynamics of the relative algorithms discrepancy
$\Delta  = |W_\mathrm{SVEA}-W_\mathrm{noSVEA}|/W_\mathrm{SVEA}$ is shown by a solid green curve.}
\label{nosvea}
\end{figure}

To test the validity of the reduced-order model discussed in \cref{sec1:5}, we perform a test, which is assumed to
violate the standard SVEA approximation, used in the averaged FDTD FEL codes. For this we consider the same
time-dependent simulation as before, but with electron current duration of 28 fs, which corresponds to only 31
wavelengths of the amplified fundamental mode. For such conditions, the slippage becomes important and should be treated
without temporal or spatial averaging. The results of the simulations with \cref{fld_eq7_nosvea} and \cref{fld_eq7} are
presented in \cref{nosvea}. The relative discrepancy in the produced radiation energy between the algorithms is also
presented with a green curve in \cref{nosvea}. The discrepancy grows with the signal amplification, however, it stays
reasonably low, and does not exceed 1.5 \%.

The one-dimensional tests were performed on a desktop computer using the 4-cores CPU Intel Xeon E5-1620. The
steady-state simulations required only 0.5 minute, while the time-dependent tests needed around 30 minutes due to the
high number of modes and macro-particles. The algorithms described in \cref{implement2:1,implement2:2} demonstrated
rather similar performance -- the non-SVEA simulation took only 4\% more time than \cref{implement2:1}. This is due to
the fact that the mentioned timing includes also particle pusher, MPI exchange, and writing the diagnostic data to the
disk.

\subsection{Three-dimensional simulations}\label{sec3.2}

An additional series of tests is performed in order to account for three-dimensional effects, such as: diffraction of
the amplified waves, focusing of the seed pulse, divergence of electron beam, and transverse structure of the even FEL
harmonics. The developed numerical code allows such study via the full three dimensional Cartesian solver, or via the
axially symmetric and antisymmetric solvers, described in \cref{sec1:1,sec1:2} respectively.

\begin{figure}[h!]\centering
\includegraphics[width=0.45\textwidth]{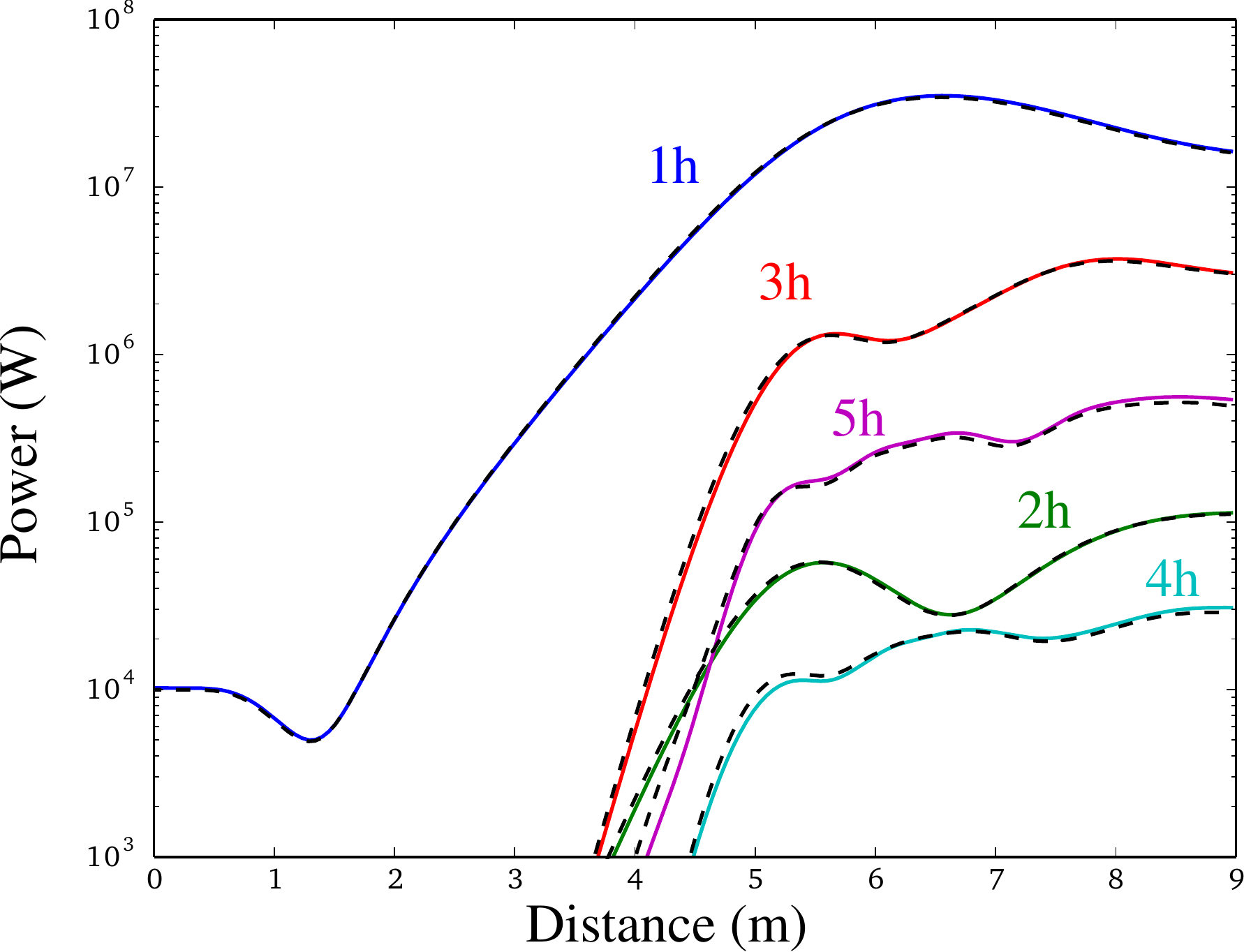}
\caption{Radiation power for the first five harmonics, as a function of the propagation distance in the steady-state
fully three-dimensional (solid lines) and axisymmetric (dashed) simulations.}
\label{3dss}
\end{figure}

For the first test, we perform a steady state modeling of the first five harmonics, $k_z = [k_s\,, 2 k_s\,, 3 k_s\,,4
k_s\,,5 k_s]$, with the fully 3D solver. The transverse size of the modeled domain is $L_x=L_y= 5$ mm, and the maximal
transverse wavenumbers are fixed to $k_x^\mathrm{max} = k_y^\mathrm{max} = 25$ mm$^{-1}$. This results in a transverse
spectral grid composed of $43\times 43$ modes. The required number of macro-particles for the steady-state simulations
can be rather low, and in this test we have used 8000 particles. The modeled growth of the radiation power is presented
in \cref{3dss} (solid curves). If compared with the data presented in \cite{giannessi:NIMA2008}, the excitation of the
third and fifth harmonics in our simulation occurs about 50 cm earlier, but agrees in the dynamics and in saturation
power values. The discrepancy may result from the difference in the undulator implementation, and in the initial
transverse distribution of the electrons in the two codes.

The fully three-dimensional simulation took 90 minutes using four 6-core Intel Xeon X5690 CPUs. Using the axisymmetric
model with the same simulation parameters, we may use only 42 transverse Bessel modes, which would drastically speed-up
the calculation. We performed a test using a combination of three solvers: one axisymmetric solver, and two
antisymmetric solvers with respect to the $x$-axis (along electron oscillations) and $y$-axis. All three solvers are
linearly independent, and may be used simultaneously for the same longitudinal modes, which allows to compare the
contributions of each solver to the total power of each FEL harmonic. The results of this test are presented by the
dashed curves in \cref{3dss} and they have a remarkable agreement with the fully three-dimensional model. In \cref{Tab1}
we introduce the fractions of the total power at the exit of the undulator modeled by each solver. As can be seen, the
odd harmonics are almost purely axisymmetric, while the even ones are antisymmetric with respect to the $y$-axis. Using
only these two solvers, and dividing the odd and even harmonics between them, this calculation can be performed in 5
minutes.

\begin{table}[h!]\centering
\caption{Fractions of the total emitted power modeled by Fourier-Bessel solvers for each of first five
harmonics}\label{Tab1}
 \begin{tabular}{c|ccc}
 Harmonic & Symmetric (\%) & $Y$-antisymmetric (\%)& $X$-antisymmetric(\%)\\
 \hline
  1h &\textbf{99.955} &0.017           & 0.028 \\
  2h &0.5             &\textbf{99.487} & 0.012 \\
  3h &\textbf{99.892} &0.052           & 0.054\\
  4h &1.458           &\textbf{98.427} & 0.115\\
  5h &\textbf{99.122} &0.677           & 0.2\\
 \end{tabular}
\end{table}

\begin{figure}[h!]\centering
\includegraphics[width=0.9\textwidth]{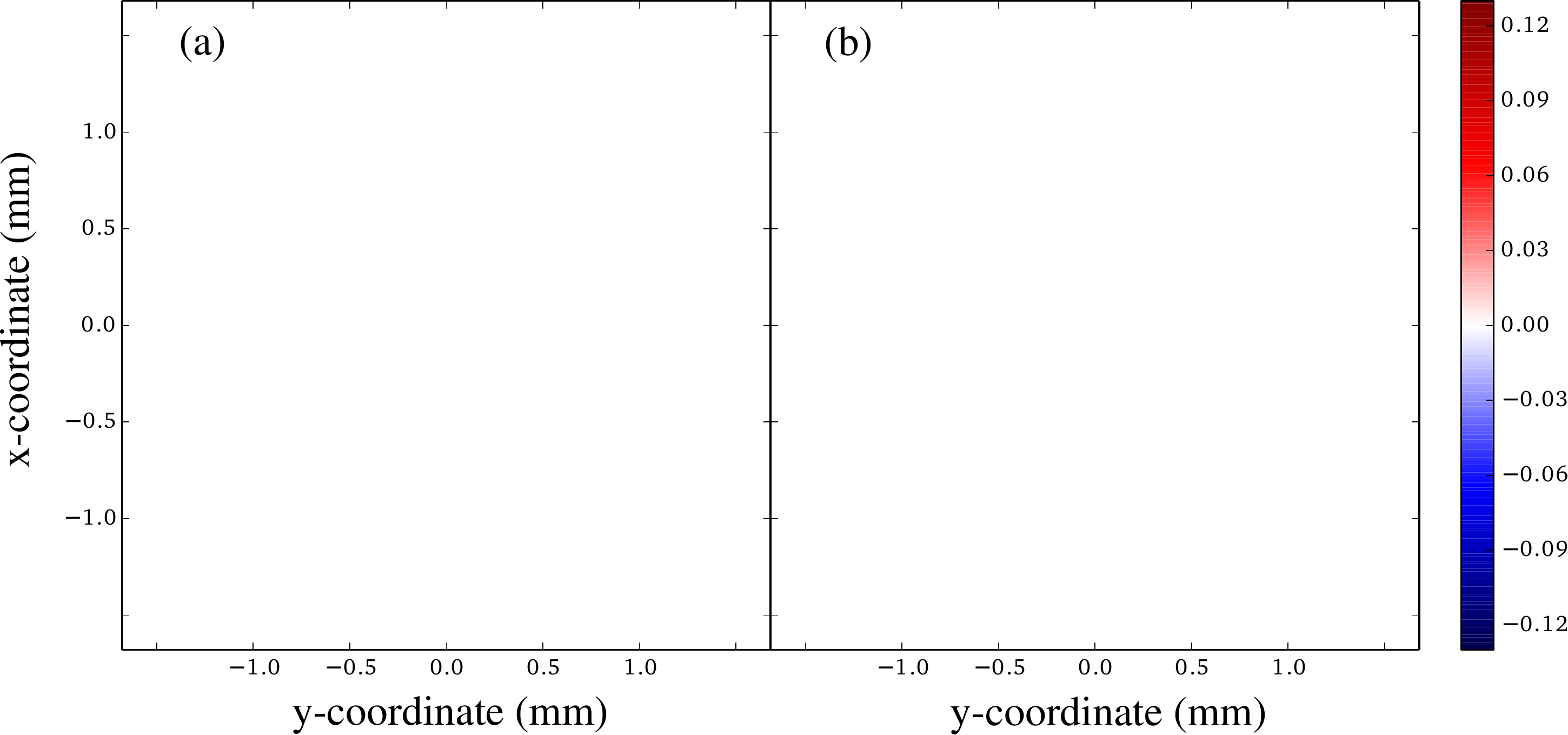}
\caption{The transverse phase structure of the second harmonic at $z=9.1$ m in the steady-state fully three-dimensional
(a) and axisymmetric (b) simulations. The value of $E_x$ field is described in the colorbar in the units of
$mc^2k_u/e$.}
\label{TEM}
\end{figure}

In \cref{TEM} we plot the phase-fronts of the second harmonic on the exit of the undulator modeled by the
three-dimensional (a) and axisymmetric (b) algorithms. The field distributions almost perfectly agree (discrepancy of
0.57 \%) except for the central part, where three-dimensional approach demonstrates slightly asymmetric features.

\begin{figure}[h!]\centering
\includegraphics[width=0.45\textwidth]{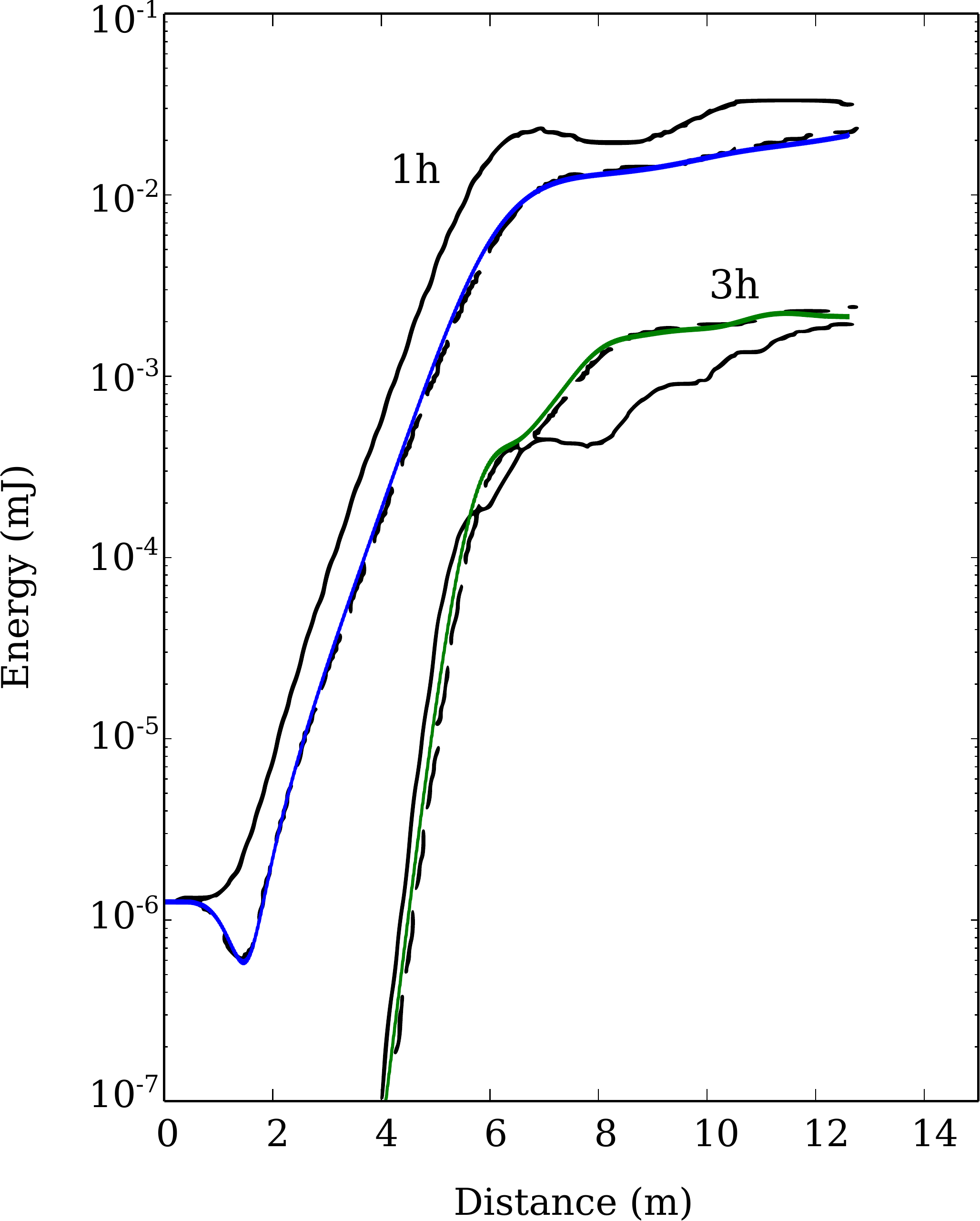}
\caption{
Energies of the first and third harmonics as a function of the propagation distance. The axisymmetric simulation
with \texttt{PlaRes} (solid colored curves) is compared with the results of PERSEO (solid black curve) and GENESIS
(dashed black curve).
} \label{compare2}
\end{figure}

In the last test, we performed a time-dependent simulation with the axisymmetric algorithms. Here, only the first and
third harmonics are modeled. The spectral domains for each harmonics are chosen to have a relative bandwidth of 1\% and
in the transverse direction the maximal wavenumber is fixed to $k_r^\mathrm{max} = 16$ mm$^{-1}$. The resulting spectral
grid for both harmonics is composed of $72\times28$ modes, and the electron beam is represented by $2\times10^5$
macro-particles. The energy growths for each harmonic are shown in \cref{compare2}a. In this figure, we compare the
results of the \texttt{Plares} simulation (colored curves) with the data produced by GENESIS (dashed black) and PERSEO
(solid black). Our results demonstrate a very good agreement with the code GENESIS, in which the three-dimensional
effects are self-consistently taken into account. On the opposite, the code PERSEO is essentially one-dimensional, and
three-dimensional effects are taken into account only approximately, by using the analytic model of M. Xie
\cite{xie:NIMA2000}.

\section{Conclusions}\label{sec4}

To conclude, we have presented a spectral time-domain method for FEL simulations, we have detailed its general
mathematical and numerical formulation. This method is used to model the resonant interaction of a short-wavelength
electromagnetic field with a relativistic particle beam. We have shown that, by representing the fields in a discrete
Fourier space, one can significantly simplify the integration of the electromagnetic equations. In doing so, we also
avoid the approximated interpolation and averaging techniques, that are associated with FTDT schemes. Instead we use the
exact projections of the particle currents onto the Fourier modes, and integrate field and motion equations resolving
the electron oscillations. The interaction of the particles with the electric and magnetic fields in real space is
described via a leapfrog integration scheme. The reduced-order field equation is derived with assumption of the
temporally slow variation of the field amplitude.

The discussed algorithms for three-dimensional and axisymmetric simulations are implemented in the code \texttt{PlaRes}.
The code is tested, and test results were compared to the ones obtained with the commonly used FEL codes. In
the test cases involving harmonic amplification in a magnetic undulator, the comparisons have
demonstrated an excellent agreement between \texttt{PlaRes} and other codes. The comparison of the models with the
second order field equation and the reduced-order SVEA equation have revealed no significant differences in the case of
a short beam.

The unaveraged algorithm for the particle motion and the general approach used in its numerical implementation provides
flexibility to the code. This allows for instance to use very realistic models for the undulators with a complex field
structure, and to consider electron beams with a complex phase-space structure. This is paramount for the concepts of
coherent radiation sources, based on the alternative drivers and undulators. For example, it has been proposed to create
a new type of X-ray sources by accelerating the electrons with a laser-wakefield accelerator \cite{fuchs:Nat2009}.
Laser-wakefield accelerators can indeed produce high peak current femtosecond beams of relativistic electrons on a
millimeter distances \cite{lundh:NatPhys2011}, but the phase-structure of these beams can be very different than that of
conventional accelerators. Presently the brightness of such sources is rather moderate as a result of the high electron
divergence. This motivates the development of more compact undulators, and a number of the alternative undulator schemes
have been proposed and explored recently \cite{petrillo:PRSTAB2008,sprangle:PRSTAB2009,andriyash:PRL2012}. In this
context, the unaveraged and generic approach of \texttt{Plares} can be beneficial, as it can be flexibility adapted to
the various proposed schemes.


\section{Acknowledgements}\label{sec5}

This work was partially supported by the European Research Council through the PARIS ERC project (Contract No. 226424)
and X-Five ERC project (Contract No. 339128). Authors would like to acknowledge Ph. Balcou, V.T. Tikhonchuk and
M.-E. Couprie for fruitful discussions.

\end{document}